# Mesoporous Thin Films as Nanoreactors for Complex Oxide Nanoparticle-based Devices

Sebastián Passanante[1,2, †], Mariano Quintero[1,2], Andrés Zelcer[3], M. Sergio Moreno[4], Diego Lionello[1,2], Daniel Vega[1], Leticia P. Granja[1,2, *]

[1] Departamento de Física de la Materia condensada, Gerencia de Investigación y Aplicaciones, Centro Atómico Constituyentes, CNEA. Av. General Paz 1499, (B1650KNA) Villa Maipú, Provincia de Buenos Aires, Argentina.
[2] Instituto de Nanociencia y Nanotecnología, Centro Atómico Constituyentes, CNEA-CONICET. Av. General Paz 1499, (B1650KNA) Villa Maipú, Provincia de Buenos Aires, Argentina.
[3] CIBION-CONICET, Godoy Cruz 2390, (C1425FQD) Ciudad de Buenos Aires, Argentina.
[4] Instituto de Nanociencia y Nanotecnología, Centro Atómico Bariloche, CNEA-CONICET. Av. Exequiel Bustillo 9500, (R8402AGP) San Carlos de Bariloche, Provincia de Río Negro, Argentina.
† Present address: Departamento de Coordinación Proyecto ICES, Centro Atómico Constituyentes, CNEA. Av. General Paz 1499, (B1650KNA) Villa Maipú, Provincia de Buenos Aires, Argentina.

* Corresponding author: leticiagranja@integra.cnea.gob.ar / granja.l@gmail.com

*We combine for the first time the properties of ordered mesoporous thin films and complex oxide nanoparticles in the design of new heterostructures, taking advantage of the accessible tridimensional pores network. In this work, we demonstrate the feasibility of synthesizing $La_{0.88}Sr_{0.12}MnO_3$ inside the pores of a mesoporous $SiO_2$ thin film, using pulsed laser deposition. In order to understand the filling process, a set of samples were deposited for three different deposition times, on mesoporous and non-mesoporous $SiO_2$ substrates. Their structural, magnetic, magnetocaloric and electrical transport properties were studied. All the results evidence the presence of the manganite compound inside the pores, which was confirmed by cross-section elemental mapping. X-ray reflectometry shows that it is possible to control the filling of the pores, keeping some accessible porosity. The magnetic behavior suggests the presence of weakly interacting ferromagnetic nanoparticles inside the pores. We provide here a successful strategy for the fabrication of complex oxide nanoparticles arrays with highly controlled size and ordering. Their easy incorporation into micro and nanofabrication procedures unveils direct implications in the field of interfaces and nanoparticle devices as diverse as energy conversion systems, solid oxide fuel cells, spintronics and neuromorphic memristor networks.*

## 1. INTRODUCTION

In the last decades, a lot of research has been focused on developing fabrication methods and applications of both nanoparticles (NPs) and thin films independently. Within this context, mesoporous oxide thin films have emerged as a successful alternative to integrate metallic NPs into thin films, generating nanostructures compatible with microfabrication techniques [1]. Additionally, the incorporation of some simple oxide and semiconductor NPs have been also reported [2]. Also, several chemical methods have been intended to incorporate halide perovskites nanocrystals into mesoporous matrices for optoelectronics applications [3]. However, the incorporation of complex oxide NPs into micro and nanodevices is still a challenge, due to their structural and compositional complexity.

Ordered mesoporous oxides, usually obtained through a combination of the sol-gel method and the self-assembly of surfactants, are materials with high surface area, controlled porosity, and ordered array of pores that can be tuned in the 2–20 nm diameter range [4], [5]. From this perspective, they are currently used in optical applications [6], [7], catalysis [8], [9], drug delivery [10], sensors [11], gas separation [12], [13], and many other fields related to nanotechnology [14], [15], [16]. When these materials are processed as thin films, the pores can act as a template for the synthesis of nanoparticle arrays. Thus, the diameter of the pores controls the size of the NPs, while the distance between the pores can tune their interaction. The quality of the mesoporous films, which present a very low rugosity, makes them ideal candidates to be incorporated as a process stage into microdevice fabrication. Moreover, the pores can be filled with different species (solvents, organic molecules, nanoparticles, etc.) to adjust and improve the properties of the materials. Currently, there is a great progress in the synthesis of metallic NPs within nanopores, and many different methods have been developed [1], where the metallic NPs obtained are composed of single elements (being Au, Ag, and Cu the most usual) [2], [17].

Mixed valence oxides of Mn (manganites) are complex compounds with a delicate stoichiometric and structural equilibrium which gives place to a wide range of fascinating properties, arising from the strong interplay between spin, charge, orbital, and lattice degrees of freedom [18]. Nowadays, this multifunctionality promotes manganites as excellent candidates for very different applications [19], including spintronics [20], [21], magnetocaloric [22], memristive devices [23], magnetic hyperthermia therapy [24], [25] and mixed conduction cathodes for solid oxide fuel cells [26], [27]. This fact has driven the effort to develop a great variety of synthesis routes in order to get very different compounds and morphologies, from single crystals and ceramics [18] to multiple nanostructures (i.e. epitaxial thin films [28], nanotubes/nanowires [29], [30] and nanoparticles [31]). Indeed, advances in manganite mesoporous thin films, following an evaporation-induced self-assembly synthesis, have been also reported [32].

We propose here to exploit mesoporous oxide thin films as substrate-confined nanoreactors [33] for the synthesis of manganite nanoparticle arrays. With this aim, we have successfully sorted out the synthesis obstacles just using the pulsed laser deposition (PLD) technique. Unlike usual chemical filling processes, PLD prevents the pores occlusion due to synthesis wastes [3]. In PLD, a target of the desired compound is ablated by a pulsed laser, creating a highly directed plasma of the constituent elements in the shape of a plume, which is steered toward the substrate where the deposition takes place [34], reproducing the stoichiometry of the target [35].

In this work, we have deposited $La_{0.88}Sr_{0.12}MnO_3$ (LSMO) by PLD on mesoporous $SiO_2$ films (MSF). The procedure is schematized in Figure 1. The MSF had been synthesized using sol-gel and evaporation induced self-assembly (EISA) techniques [5], on $SiO_2$/Si substrates. Bulk LSMO has a paramagnetic (*PM*) to ferromagnetic (*FM*) transition temperature ($T_C$) at 290 K [36]. However, the magnetic behavior of NPs systems strongly depends on geometrical parameters, mainly the size and shape of the NPs and the distance between them [37]. Moreover, the complexity of the LSMO phase diagram region [38] enhances these effects. LSMO was deposited on mesoporous and non-mesoporous $SiO_2$, varying the amount of material ablated, to systematically study the effect of the porous constriction on the LSMO properties. Silica was chosen because it is unreactive and remains amorphous up to very high temperatures (>1000 °C), so it is not expected to interact chemically with manganite, nor any grain growth will modify the mesostructure during the LSMO deposition, performed at 850 °C.

We present here the structural, magnetic, and electrical characterization of the heterostructures prepared by combining PLD and ordered mesoporous thin films. The results confirm the presence of LSMO inside the pores of the mesoporous silica films. Moreover, as manganites are good candidates for magnetocaloric effect (MCE) applications [39], we also explore the MCE properties of LSMO deposited on mesoporous and non-mesoporous $SiO_2$ systems. These results support the application of manganite NPs to cool / heat mesoporous oxide layers for microdevice applications.

In short, our work widens the applications of complex oxides nanoparticles, which were synthesized for the first time in a well-defined geometrical network. Regarding that the quality of these heterostructures makes them suitable for microfabrication process, specific applications for spintronic, solid oxide fuel cells (SOFCs) and magnetocaloric devices are proposed here.

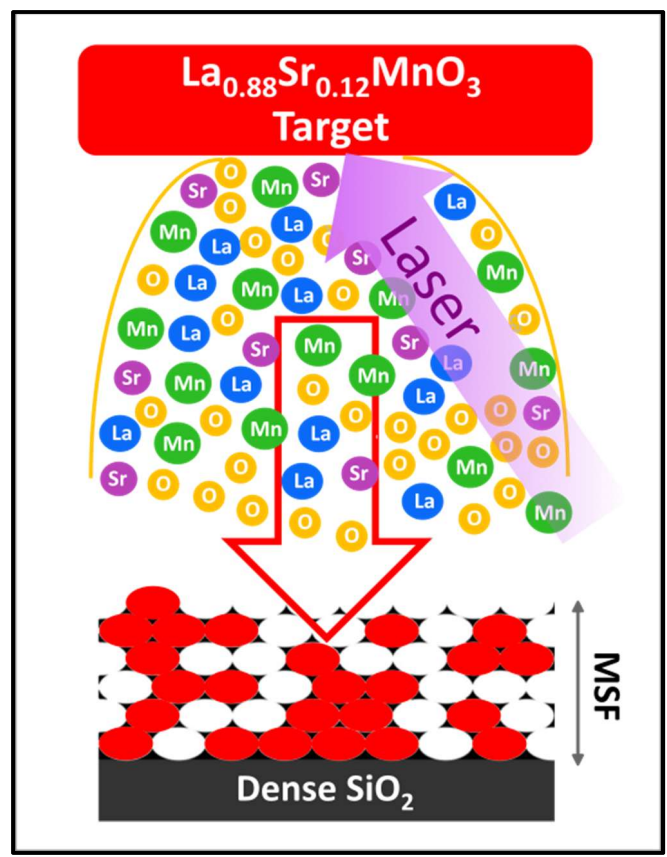

Figure 1: Scheme of the PLD process onto a mesoporous thin film. LSMO is separated into its constituent elements and recombined at the mesoporous substrate.

## 2. RESULTS AND DISCUSSION

Mesoporous $SiO_2$ films were deposited by dip coating on a commercial Si substrate with 1 μm of dense and amorphous thermal oxide ($SiO_2$/Si substrates), using a sol-gel route in combination with the EISA strategy [40]. The templating agent used was commercial Pluronic F127, which was eliminated after a thermal treatment at 350 °C (see Experimental Section for the synthesis details). Dip coating method guarantees large homogenous areas of high-quality films. The obtained mesoporous $SiO_2$ thin film has an *Im3m* cubic mesostructure, with a thickness of (98 ± 5) nm and pore size of 8 nm (See Figure 2.a). After deposition, it was cut into several pieces in order to use the same MSF for all the samples studied in this work.

$La_{0.88}Sr_{0.12}MnO_3$ (LSMO) was deposited by PLD, using three different deposition time ($t$) on the MSF described above, in order to obtain samples with different amount of LSMO. Like $t$ determines the expected film thickness ($d$), samples are named using the expected thickness $d$ for the deposited LSMO as LSMOd/MSF. Control samples were deposited on $SiO_2$/Si substrates using the same set of deposition times, and are correspondingly named LSMOd/$SiO_2$. Figure 2.b displays the X-ray diffraction (XRD) results for the thickest LSMO films. The angular position of the observed reflections agrees with the main peaks of the orthorhombic Pnma structure [41], suggesting that the LSMO composition is conserved after deposition. The complete set of XRD patterns for all the films is displayed at the Supporting Information (SI).

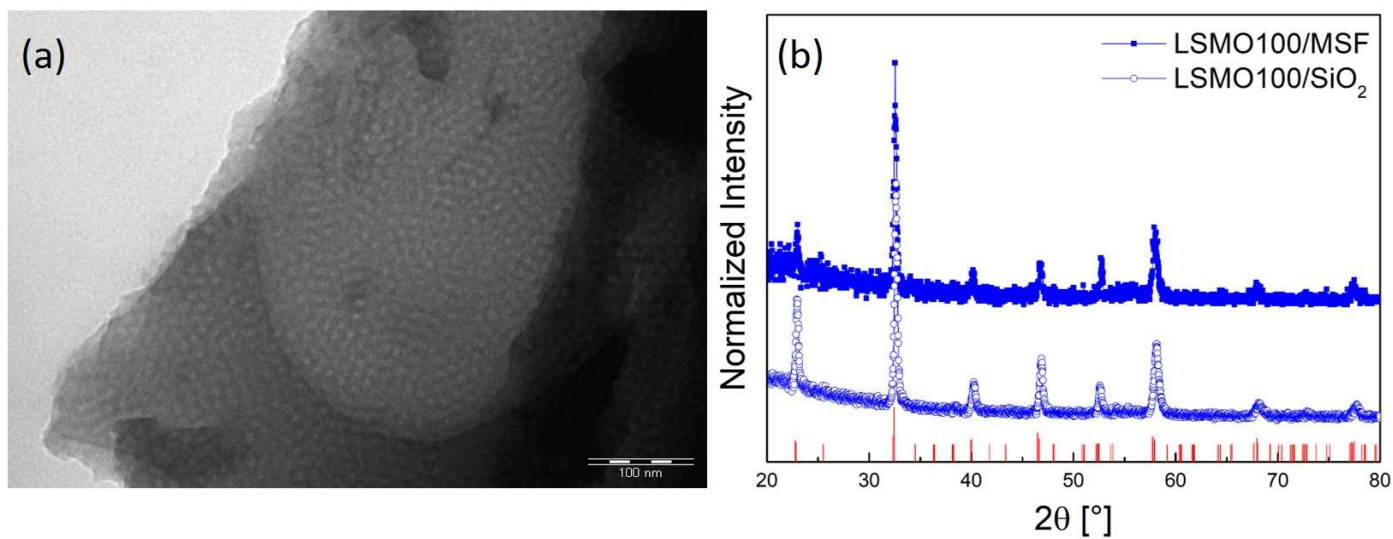

Figure 2: (a) TEM image of an MSF flake. (b) XRD patterns obtained for LSMO100/$SiO_2$ and LSMO100/MSF, compared with the orthorhombic Pnma reflections for LSMO [41]. XRD measurements for all the samples can be found in Figure S1.

Scanning electron microscopy (SEM) characterization of the LSMOd/MSF samples is shown in Figure 3. The surface of the MSF displays the typical self-assembled silica mesostructure obtained using the Pluronic F127 molding agent (Figure 3.a) [42]. The morphology of LSMO films deposited on MSF displays a strong dependence on the deposition time. LSMO100/MSF (Figure 3.b), which is the longest $t$ sample (300 s), presents the polycrystalline film growth usually observed for films deposited on dense $SiO_2$/Si substrates (see Figure S7), with an average grain size of ~35 nm [43]. Reducing the deposition time to $t$ = 60 s (LSMO20/MSF), yields a similar polycrystalline film growth over the mesoporous surface (Figure 3.c). However, Figure 3.d shows that LSMO5/MSF, the sample with the shortest $t$ (15 s), presents very similar features to the MSF sample (Figure 3.a), without any continuous polycrystalline LSMO film on its surface. This suggests that the manganite was deposited inside the pores of the MSF. Transmission electron microscopy (TEM) confirms that the mesostructure remains ordered for all the samples after PLD procedure, even under the polycrystalline LSMO layer. More details on SEM and TEM characterization of MSF and LSMOd/MSF samples and SEM characterization of control samples can be found at the SI.

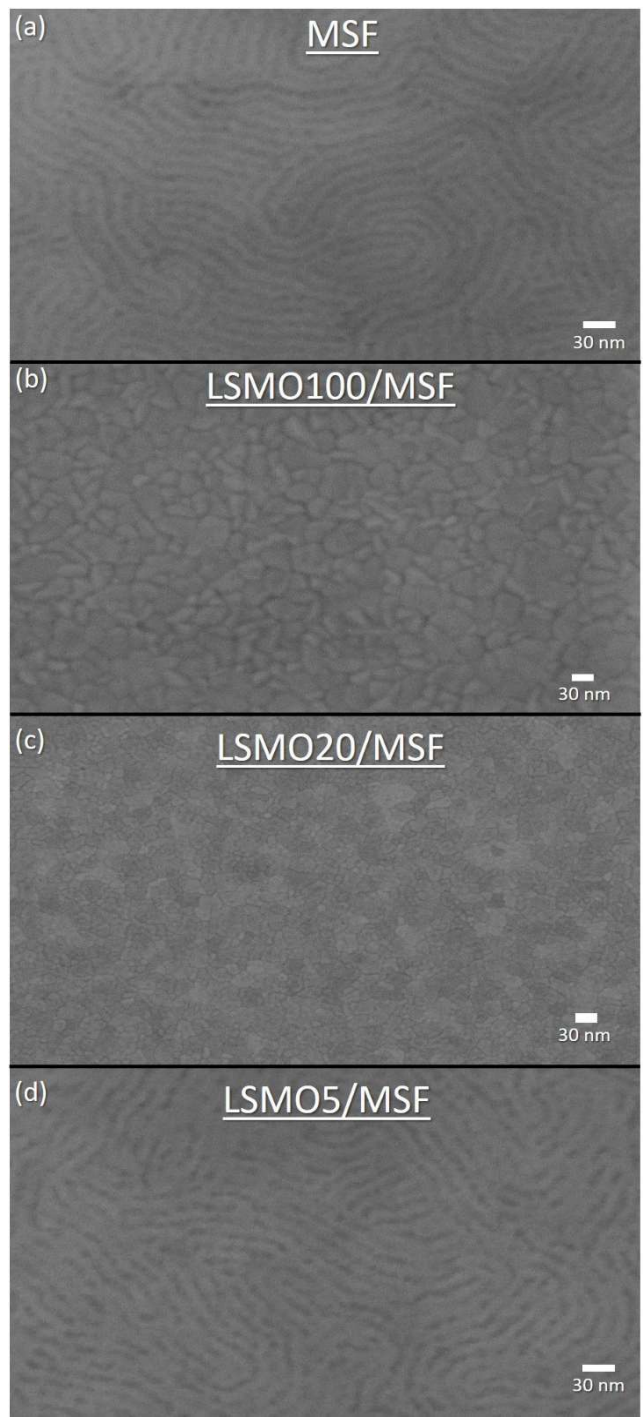

Figure 3: SEM images of the surface of: (a) MSF, showing the mesoporous pattern; (b) LSMO100/MSF and (c) LSMO20/MSF, with a polycrystalline growth on the surface of MSF; (d) LSMO5/MSF, showing a similar pattern to MSF instead of a polycrystalline film.

Zero-loss energy filtered electron microscopy (HRTEM) images of the cross-section of LSMO5/MSF and LSMO20/MSF lamellas are shown in Figure 4. In both cases, a clear contrast between the different layers is observed and the porous structure is clearly distinguished in the MSF layer. A careful inspection and analysis of the TEM images were performed to get deeper information about the crystalline structure of both: the continuous nanocrystalline LSMO thin film and the nanoparticles synthetized within the pores of MSF (see section IV of the SI). Thus, the lattice planes observed at the LSMO top layer for LSMO20/MSF (Figure S11), confirmed the orthorhombic Pnma manganite phase suggested by XRD (Figure 2.b). Even more, despite the low crystallinity of the nanoparticles inside the pores, we found some lattice planes in nanoparticles of LSMO5/MSF flakes which agree with the (122) and (031) reflections of the orthorhombic structure, bearing out its composition (Figure S12).

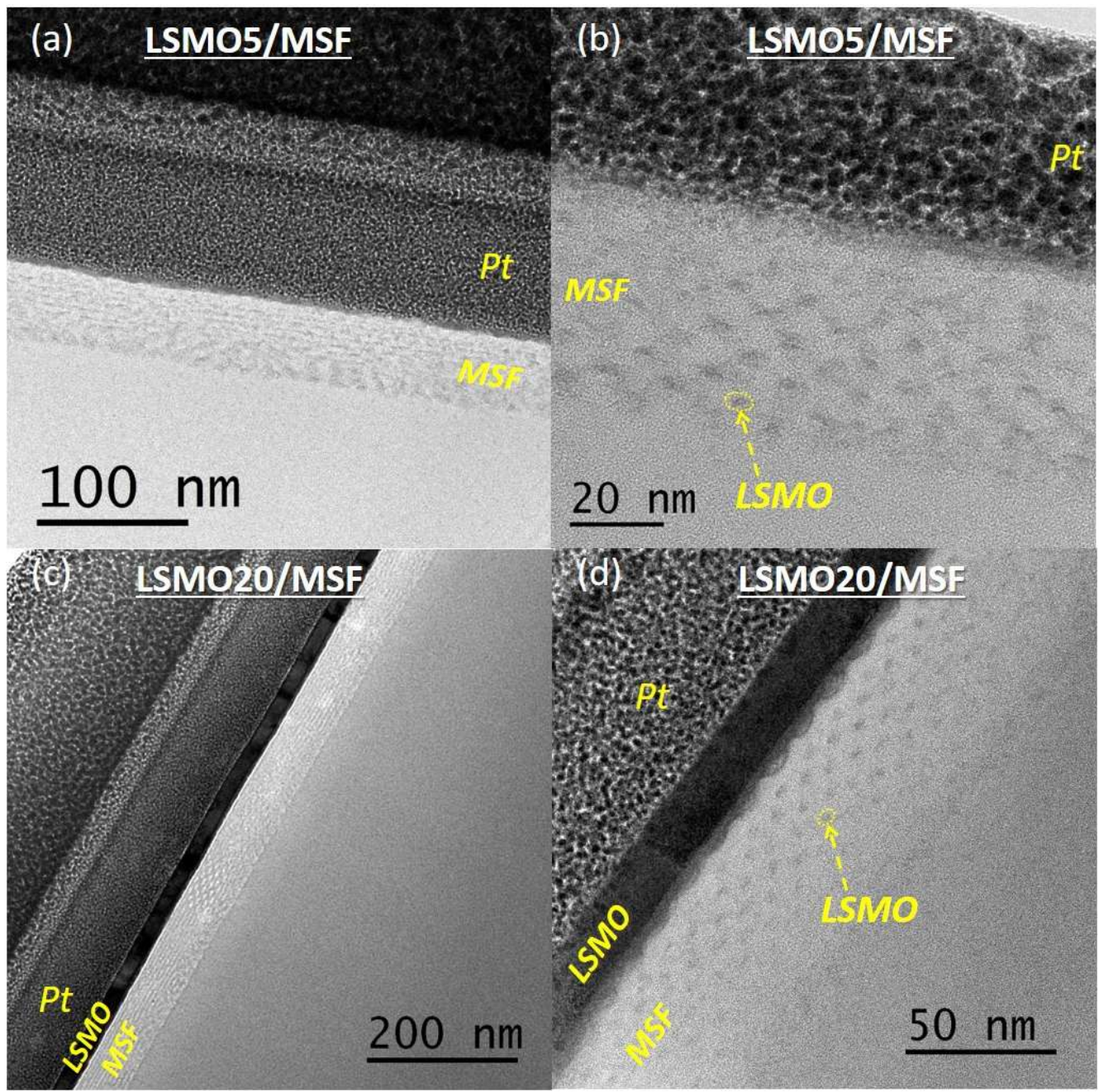

Figure 4: Zero-loss filtered TEM images in cross-section geometry of: (a), (b) LSMO5/MSF and (c), (d) LSMO20/MSF.

We used Electron energy-loss (EEL) spectroscopy to better understand the morphology and composition of the obtained systems, particularly in the region of the mesoporous layer remarked in Figure 5.a. The EEL spectrum acquired (Figure 5.d) shows contributions from the oxygen-K edge, Mn-$L_{2,3}$, and La-$M_{4,5}$ edges. As expected, the oxygen edge signal dominates the spectrum. Therefore, there is a partial overlap of the Mn–$L_{2,3}$ edge and the oxygen contribution which imposes restrictions for Mn mapping. Because of this reason, the jump-ratio method was chosen to map La and Mn (It is not possible to map Sr because its $L_{2,3}$ edge overlaps with the K edge of Si). Elemental maps for La and Mn are shown in Figures 5.b and 5.c. Both maps show a mottled contrast at the spatial scale of the mesoporous layer which agrees with the size and array of the pores (about 8 nm), indicating that both elements are present inside the MSF layer. It is remarkable that both La and Mn are homogenously distributed from the lower interface with the dense $SiO_2$ layer to the upper interface with the LSMO layer.

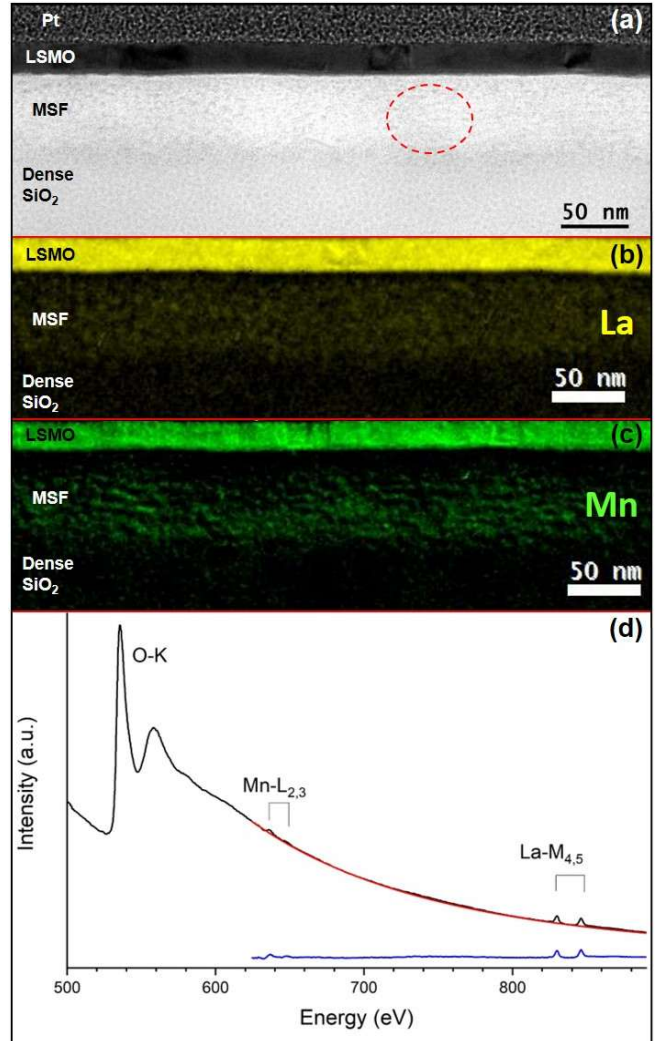

Figure 5: (a) Bright-field TEM image of LSMO20/MSF in cross-section geometry where EEL spectrum was acquired (indicated by the red circle). (b) La (yellow) and (c) Mn (green) elemental maps corresponding to the cross-section region selected in Figure 5.a. (d) EEL spectrum of the mesoporous region indicated by the red circle in Figure 5.a. The O-K, Mn-$L_{2,3}$, and La-$M_{4,5}$ edges are visible. (The blue curve is the result of the background (red) subtraction from the spectrum (black)).

Very valuable information about the morphology and composition of thin films can be deduced from X-ray reflectometry (XRR) measurements [44]. The critical angle of reflection ($\theta_c$) is directly related to the electronic density ($\delta$) as $\delta = \frac{\pi \theta_c^2}{\lambda^2 r_e}$ (Eq. 1), where $\lambda$ is the X-rays wavelength and $r_e$ is the classical electron radius. After X-rays penetrate the film, an interference pattern appears whose maximum and minimum locations are determined as $\theta_{max/min}^2 = \theta_c^2 + (m + \Delta m)^2 \cdot \frac{\lambda^2}{4d^2}$ (Eq. 2), where $m$ is the order of the maximum (minimum) fringe at $\theta_{max}$ ($\theta_{min}$) and $d$ is the film thickness.

Figure 6 displays the XRR measurements for the LSMOd/MSF set of samples (Figure 6.a) and the comparisons with their respective LSMOd/SiO$_2$ samples (Figures 6.b - 6.d). Table 1 presents the critical angle and the thickness of the films extracted from these XRR measurements. For LSMO100/MSF and LSMO20/MSF, $\theta_c$ agrees with those obtained for all LSMOd/SiO$_2$ samples (see Table 1), suggesting that a continuous top layer of LSMO is present in LSMO20/MSF and LSMO100/MSF, unlike LSMO5/MSF, whose $\theta_c$ is just slightly larger than the value observed for bare MSF.

Morphological differences are clearly manifested in the XRR curves once the critical angle is overcome. For the thickest manganite samples, *d* = 100 (Figure 6.b), interference oscillations are dominated by the LSMO layer. In contrast, Figures 6.c and 6.d show that the oscillation patterns for LSMO20/MSF and LSMO5/MSF are more complex than those obtained for a single LSMO layer deposited on SiO$_2$. The effects of the presence of the MSF layer are superimposed on the modulation of a thinner LSMO top layer in the XRR curves.

Particularly for LSMO5/MSF (Figure 6.d), two different frequencies are present in the XRR curve. On one hand, the lowest frequency modulates the amplitude following the LSMO5/SiO$_2$ oscillations. However, regarding that in the electron microscopy images no continuous LSMO layer is observed (Figures 3.d and 4.b), just a porous thin LSMO layer could be present, mimicking the superficial porous texture of the MSF. On the other hand, the highest frequency clearly emulates the maximums and minimums pattern obtained for the MSF sample (Inset of Figure 6.a). These fringes render a thickness of (75 ± 5) nm, in agreement with the thickness observed for MSF in the cross-section images of LSMO5/MSF and LSMO20/MSF (Figure 4), which is smaller than the one obtained from MSF (Table 1), suggesting that the MSF film suffered a thickness contraction due to the 850 °C heating temperature during the PLD procedure. Moreover, a peak is observed at *2θ* = 1.21° for MSF and *2θ* = 1.45° for LSMO5/MSF, which corresponds to the first Bragg peak of the diffraction produced by the planes of pores parallel to the substrate [45], [46]. The difference in the position of this peak between both samples is another evidence of the vertical contraction of the mesostructure. This peak is also present in the LSMO20/MSF XRR pattern, but it is split due to the overlap of the interference within the LSMO and the MSF layers (Fig. 6.c). For LSMO100/MSF, the peak becomes negligible in the XRR curve (Fig. 6.b), because the LSMO layer is thick and dense enough to mask the interference. In the case of the LSMOd/SiO$_2$ samples, no peak appears due to the absence of the MSF bottom layer. The effects of the deposition temperature and atmospherical conditions are studied in the Sections ii, iii and vi of the SI, confirming that the mesoporous structure remains ordered and accessible despite thickness contraction.

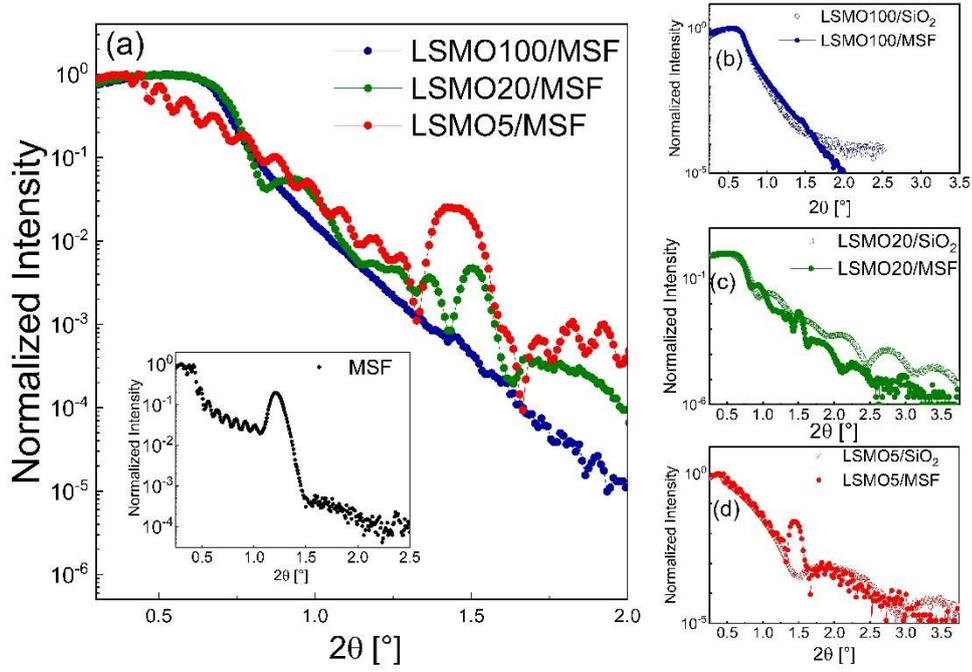

Figure 6: (a) XRR measurements for LSMOd/MSF samples performed in low relative humidity conditions (*RH* < 15%). Inset: XRR of the MSF film used for the LSMO deposition. XRR results comparing (b) LSMO100/MSF and LSMO100/SiO$_2$, (c) LSMO20/MSF and LSMO20/SiO$_2$ and (d) LSMO5/MSF and LSMO5/SiO$_2$.

| Sample | *PLD Substrate* | *t* [s] | Thickness [nm] | θ$_c$ [°] |
|---|---|---|---|---|
| **LSMO5/MSF** | MSF/SiO$_2$/Si | 15 | --- (*) | 0.200 ± 0.004 |
| **LSMO20/MSF** | MSF/SiO$_2$/Si | 60 | 20 ± 1 | 0.358 ± 0.004 |
| **LSMO100/MSF** | MSF/SiO$_2$/Si | 300 | 85 ± 3 | 0.350 ± 0.004 |
| **LSMO5/SiO$_2$** | SiO$_2$/Si | 15 | 5 ± 1 | 0.330 ± 0.030 |
| **LSMO20/SiO$_2$** | SiO$_2$/Si | 60 | 14 ± 2 | 0.349 ± 0.004 |
| **LSMO100/SiO$_2$** | SiO$_2$/Si | 300 | 109 ± 8 | 0.350 ± 0.004 |
| **MSF** | --- | ---- | 98 ± 5 | 0.170 ± 0.004 |

Table 1: Substrate and deposition time *t* used during the PLD procedure, film thickness and critical angle θ$_c$ determined by XRR. (*) It is not possible to determine the LSMO thickness by XRR.

The relation between $\delta$ and $\theta_c$ (Eq. 1) can be used to extract information about the porosity and the filling fraction (*F*) of the pores [47]. It is well known that only a fraction of the porosity is accessible to the environment because some pores will remain occluded after the surfactant calcination [48]. Therefore, it is necessary to measure the accessible porosity (*AP*) of the mesoporous thin films to quantify the filling capacity of the pores. Taking advantage of the capillary condensation inside the pores, *AP* can be estimated from the shift of $\theta_c$ in the XRR measurements performed at low relative humidity (*RH* < 15%) and at high relative humidity (*RH* > 90%) environmental conditions, which yield empty and filled pores with water vapor respectively. The filling of the pores modifies the electronic density of the film. The comparison of the XRR results from measurements performed under both *RH* conditions, can be used to deduce the accessible porosity [46] as $AP = \frac{(\delta_{wet}-\delta_{dry})}{(\delta_{H2O}-\delta_{air})} \cdot 100$ (3), where $\delta_{wet}$ and $\delta_{dry}$ are the electronic densities measured for high and low *RH* respectively, while $\delta_{H2O}$ is the electronic density for water and $\delta_{air}$ is the electronic density for the air. Therefore, *AP* represents the percentage of the total porosity that remains accessible to the external environment, being able to be filled with other species (More detailed description of this characterization can be found in Section v of the SI). The MSF film, used as bottom layer, presented $AP$ = 34% previously to be subjected to the PLD deposition conditions (See Figure S13), a value typically found on these systems [47]. It is worth noting that sample LSMO5/MSF, obtained after 15 s of deposition, still presents a measurable accessible porosity suggesting that the pores are not completely filled and a remaining porosity is still available (see Figure S14). Therefore, these findings indicate that it is possible to build a porous nanostructure, composed of a mesoporous thin film and a complex oxide, with a significant $AP$ that allows interaction with the environment. For LSMO20/MSF and LSMO100/MSF, no differences were found between either $RH$ conditions (not shown), due to the presence of the continuous and dense LSMO top layer over the MSF film.

Furthermore, in the same way as in Eq. 3, the percentage of the accessible porous volume of the MSF that was filled with LSMO for LSMO5/MSF can be estimated as: $F = 100 \cdot \frac{(\delta_{LSMO+MSF} - \delta_{MSF})}{(\delta_{LSMO} - \delta_{air})} / AP$ (4), where in this case $\delta_{LSMO+MSF}$, $\delta_{MSF}$ and $\delta_{LSMO}$ are the electronic densities for LSMO5/MSF, MSF and LSMO100/SiO$_2$, measured in low $RH$ conditions. Taking into account the observed thickness contraction of MSF during LSMO deposition, and consequently its density increment, $F$ could be approximated as an upper limit. Therefore, at most 27% of the initial $AP$ of MSF would have been filled with LSMO after 15 s of deposition (Note that $AP$ is not appreciably modified by PLD conditions, as shown in Figure S19). Thus, it would be expected that a critical deposition time exists between 15 s and 60 s, maximizing the LSMO filling fraction before covering the MSF surface with a continuous LSMO layer.

The structural characterization demonstrates that PLD technique is key, confirming the deposition of LSMO on the whole MSF accessible surfaces. Therefore, it is much more efficient than other filling methods for mesoporous thin films, where the filling procedure itself occludes the pores impeding subsequent filling [1], [2]. PLD process turns out in a bottom – top filling of the pores, followed by the formation of a continuous upper layer.

Magnetization ($M$) as a function of temperature ($T$) was measured applying a magnetic field ($H$), $H$ = 500 Oe, for field-cooled warming (FCW) and zero-field cooled warming (ZFC) conditions. $M(T)$ results were normalized with the $M$ value at the paramagnetic regime ($M_{PM}$) in order to avoid arbitrary geometrical criteria, whose discussion will be resumed later in this text. Figures 7.a – 7.f display the ZFC and FCW measurements ($M_{ZFC}(T)$ and $M_{FCW}(T)$ respectively), which present a PM-FM transition for all the samples. Table 2 shows the Curie temperature ($T_C$), obtained from their derivative. Interestingly, $T_C$ increases with lower d when LSMO is deposited on the mesoporous silica. This result is in contrast with that obtained from LSMOd/SiO$_2$, whose $T_C$ decreases when the thickness is reduced, as usual for thin films. The magnetic behavior of LSMOd/MSF samples can be associated with the LSMO within the pores, whose magnetic contribution becomes more appreciable with the reduction of the thickness of the LSMO top layer. Particularly the $T_C$ = 290 K obtained for LSMO5/MSF is consistent with the $T_C$ reported in LSMO nanoparticles [49], [50].

Comparing $M(T)$ results for the films deposited on MSF at low $T$ (inset of Figures 7.a, 7.c, and 7.e), a clear difference between ZFC and FCW curves appears. Note that the $H$ = 500 Oe value was chosen to be greater than the coercive field for all the samples in the whole range of $T$ explored, in order to separate properly the magnetic effects. The behavior of $M_{ZFC}(T)$ at low temperatures suggests that part of the material was frozen in a magnetically disordered state ($H$ = 0). On warming with $H$ = 500 Oe, $M_{ZFC}(T)$ remains blocked until $T$ is high enough to unblock the system, which is evidenced in the convergence of $M_{ZFC}(T)$ and $M_{FCW}(T)$ to the same $M(T)$ values. This behavior is enhanced with the reduction of the LSMO top layer, suggesting that it proceeds from the inner of the MSF layer, being supported by the one observed for weakly interacting magnetic nanoparticles [37]. By contrast, in the case of LSMOd/SiO$_2$, no differences between $M_{ZFC}(T)$ and $M_{FCW}(T)$ curves are observed beyond a slight difference due to the magnetic dead layer [51], [52]. Therefore, the magnetic behavior of LSMOd/MSF supplies an indirect confirmation of the confinement of LSMO NPs inside the pores of MSF.

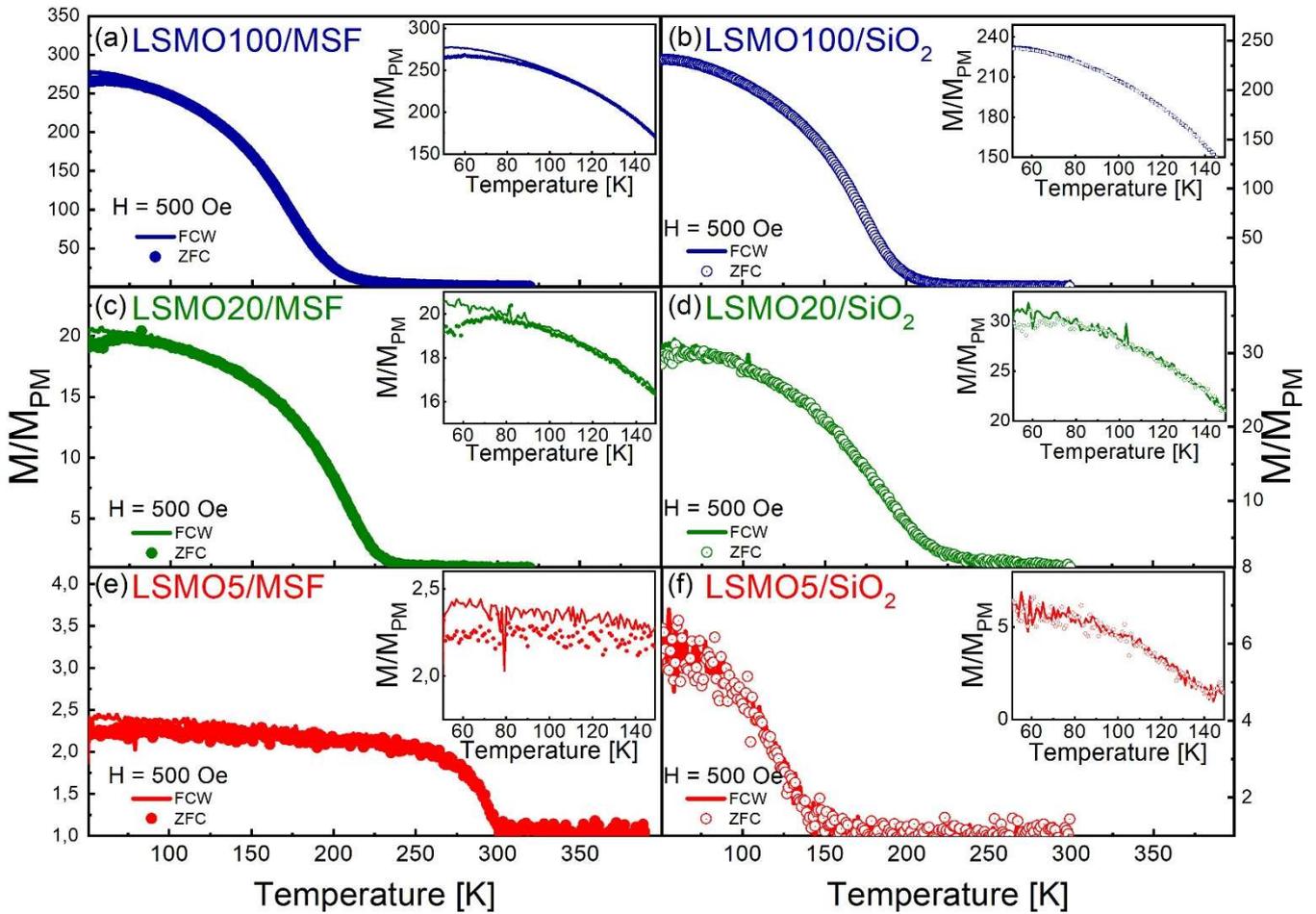

Figures 7.a – 7.f: M(T) of LSMOd/MSF and LSMOd/SiO$_2$ samples, measured with H = 500 Oe in field-cooled warming (FCW) and the zero-field cooled (ZFC) conditions. The insets show a zoom of the low temperature part of each curve.

Figures 8.a - 8.c present the $M(H)$ curves measured at 50 K. The measurements were carried out between -30000 Oe and 30000 Oe (see Figure S21 for full range data). $M$ was normalized considering the volume of the upper LSMO film ($V_{LSMO}$) for each sample, except LSMO5/MSF which is discussed below. All the samples display the typical FM behavior, although differences between the LSMOd/MSF and LSMOd/SiO$_2$ systems can be clearly distinguished. The presence of the pores seems to increase the coercive field ($H_c$), rendering LSMOd/MSF magnetically harder than LSMOd/SiO$_2$ (Figure S22 displays that this behavior holds along the full T range explored). In most of the cases, $M(H)$ reached a saturated magnetic state ($M_{sat}$) at $H \ll 30000$ Oe. Table 2 summarizes the main results extracted from the data presented in Figures 7 and 8. For LSMO100/MSF and LSMO20/MSF, their $M_{sat}$ values are comparable. The same occurs for LSMO100/SiO$_2$ and LSMO20/SiO$_2$. However, it is remarkable that $M_{sat}$ is 0.5 μ$_B$ / f.u. greater for the films grown on MSF. These results suggest that there is a magnetic contribution not considered in the geometrical normalization, which would come from the LSMO inside the pores. The situation is very different for the thinnest manganite films (Figure 8.c). On the one hand, the behavior of LSMO5/SiO$_2$ is in accordance with the one usually reported for very thin films [53], where the influences of the magnetic frustration at the interfaces compete with $H$, resulting in a smaller $M_{sat}$. But on the other hand, the magnetic response of LSMO5/MSF is qualitatively different to the rest: the hysteresis range is wider in $H$ and the saturation is not reached even at 30000 Oe. This $M(H)$ behavior is consistent with the presence of weakly interacting single domain nanoparticles, whose possible scenarios, including superparamagnetic and glassy low temperature regimes, have been thoroughly discussed in the literature for manganite nanoparticles [51], [54], [55].

As there is no continuous LSMO top layer in LSMO5/MSF, the thickness of the LSMO could not be determined by XRR, SEM nor cross-section HRTEM images of the LSMO5/MSF lamella (see Figure 4.b). This fact hindered $M$ normalization in terms of geometrical parameters. Thus, the volume of material deposited was estimated just considering $M_{sat} \leq 3.88$ $\mu_B$/f.u., which is the theoretical saturation value obtained if all the Mn atoms of the LSMO are oriented parallel to $H$. Considering a continuous film of the area of this sample, it was found that the thickness needed to achieve $M_{sat} = 3.88$ $\mu_B$/f.u would be 5.3 nm, which is very close to the expected value of 5 nm estimated from the deposition rate (see Table 1). If we perform geometrical considerations to calculate the filling fraction of the accessible porosity of MSF for the volume of LSMO deposited ($V_{LSMO}$ = sample area × 5.3 nm), we obtain that this $V_{LSMO}$ would fill ~24% of the accessible mesoporous structure (See Section v of the SI for the details of this calculation). Note that this result is amazingly consistent with the filling fraction $F \sim 27\%$ obtained with Eq. 4 from the XRR measurements. Furthermore, from the magnetic response of LSMO5/MSF, it can be confirmed that the volume of LSMO deposited is not enough to fill the available porous volume in the MSF.

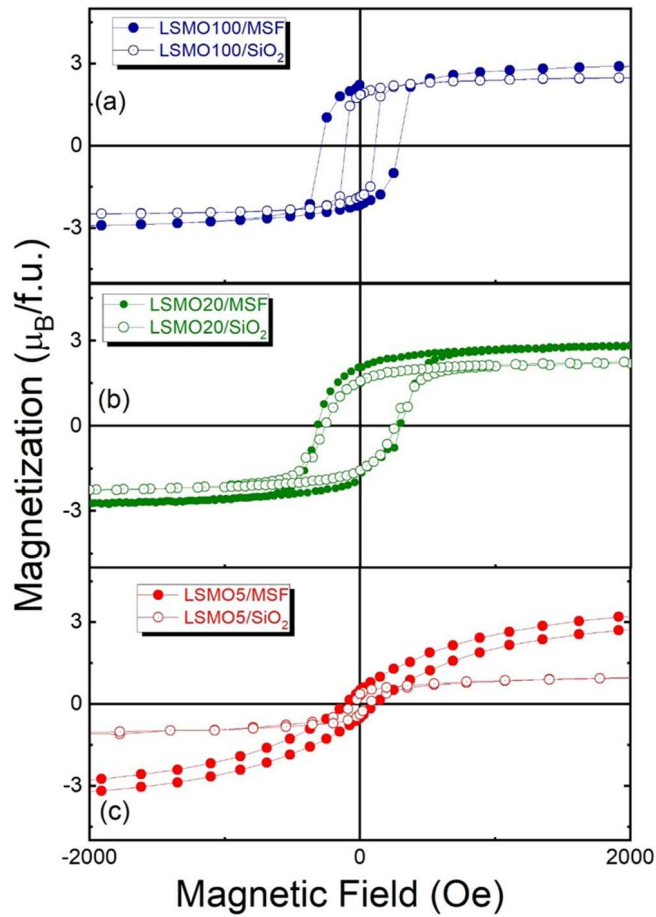

Figure 8: Magnetization vs applied magnetic field measured at 50 K. Comparison between (a) LSMO100/MSF and LSMO100/SiO$_2$, (b) LSMO20/MSF and LSMO20/SiO$_2$ and (c) LSMO5/MSF

and LSMO5/SiO$_2$, in the interval -2000 Oe < H < 2000 Oe, is shown to display the hysteresis details. (Full range curves, with -30000 Oe ≤ H ≤ 30000 Oe, are shown in Figure S21)

| Sample | $T_c$ [K] | $H_c$ [Oe] | $M_{sat}$ [μ$_B$ / f.u.] |
|---|---|---|---|
| **LSMO5/MSF** | 290 ± 5 | 113 ± 1 | (*) |
| **LSMO20/MSF** | 205 ± 5 | 305 ± 9 | 2.9 ± 0.1 |
| **LSMO100/MSF** | 175 ± 5 | 287 ± 2 | 3.1 ± 0.1 |
| **LSMO5/SiO$_2$** | 120 ± 5 | 60 ± 7 | 1.0 ± 0.1 |
| **LSMO20/SiO$_2$** | 180 ± 5 | 258 ± 2 | 2.4 ± 0.1 |
| **LSMO100/SiO$_2$** | 175 ± 5 | 110 ± 1 | 2.5 ± 0.1 |

Table 2: Summary of the main results obtained from the magnetic measurements displayed in Figures 7 and 8. (*) As the main volume of the LSMO material was deposited inside the mesoporous film, it was not possible to use the size of the LSMO film to normalize $M$ nor $M_{sat}$ was reached in this case.

As it is known, manganites are promising materials for magnetocaloric applications, and particularly LSMO presents magnetocaloric effect (MCE) [22]. Moreover, nanostructuration and morphology effects were widely explored for improving the design of MCE devices [56], [57]. The MCE is the increment (decrease) in the temperature of the sample, detected when an external magnetic field is applied (removed) adiabatically. This effect is stronger around T$_C$. Therefore, the incorporation of nanoparticles with MCE into the mesoporous matrix opens the possibility to locally control the temperature of the surrounding film by applying an external magnetic field. A usual procedure to characterize the MCE consists of measuring $M(H)$ curves at different temperatures. Then the isothermal magnetic entropy change ($\Delta S_M$) is calculated using the expression $\Delta S_M = \frac{1}{\Delta T}\int_0^H (M(T+\Delta T, H) - M(T, H))dH$, where $\Delta S_M$ is related to the adiabatic temperature change as $\Delta T_{ad} = -\frac{T}{C}\Delta S_M$ [22]. Figure 9 shows $-\Delta S_M(T)$ calculated at H = 6000 Oe, for both LSMOd/MSF and LSMOd/SiO$_2$ systems (the set of M(H) measurements employed to carry out this analysis is displayed in the Figures S23 – S28). In all the cases, a well-defined peak centered at T$_C$ was obtained, which is the maximum change in the magnetic entropy ($|\Delta S_{Max}|$) and indicates how strong the MCE is around $T_C$. Another important parameter is the width in temperature of the peak, defined as the temperature interval at full-width half maximum ($\delta T_{FWHM}$), which represents the temperature range where the MCE is stronger. Table 3 summarizes these results extracted from Figure 9.

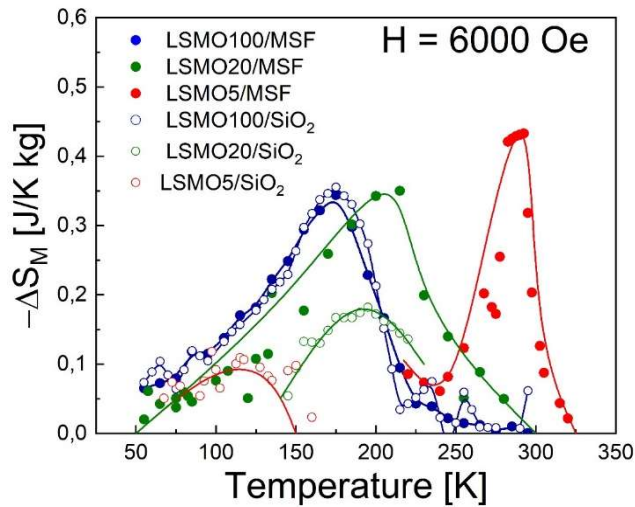

Figure 9: Isothermal entropy change $\Delta S_M$ as a function of the temperature, calculated for $H$ = 6000 Oe (Lines are guides for the eye).

| Sample | $|\Delta S_{Max}|$ [J/kg k] | $\delta T_{FWHM}$ [K] |
|---|---|---|
| **LSMO5/MSF** | 0.45 | 21 |
| **LSMO20/MSF** | 0.35 | 80 |
| **LSMO100/MSF** | 0.34 | 82 |
| **LSMO5/SiO$_2$** | 0.11 | 89 |
| **LSMO20/SiO$_2$** | 0.18 | 96 |
| **LSMO100/SiO$_2$** | 0.35 | 80 |

Table 3: Magnetocaloric effect characteristic parameters for LSMOd/MSF and LSMOd/SiO$_2$ samples.

It is interesting to note that LSMOd/SiO$_2$ systems show a decrease of the MCE with the reduction of $d$ (see in Figure 9). Moreover, there are no significant differences in $-\Delta S_M (T)$ between LSMO100/MSF and LSMO100/SiO$_2$. The LSMO20/MSF curve is similar to the d = 100 case, but presents a broader $\delta T_{FWHM}$, without any reduction of $|\Delta S_{Max}|$. However, it is worthy to note that LSMO5/MSF presents a completely different behavior from the other samples, displaying a higher maximum and a narrower $\delta T_{FWHM}$. As no upper continuous LSMO layer is present in LSMO5/MSF, the MCE observed should be mainly attributed to the LSMO deposited inside the MSF pores. Indeed, part of the observed MCE response of LSMO20/MSF should also come from the LSMO deposited inside the pores. In contrast, the top LSMO layer of LSMO100/MSF would be thick enough to completely dominate the MCE response, masking any effect coming from nanoparticles of LSMO inside the pores. Thus, these results are promising for the design of MCE devices based on heterostructures including mesoporous thin films, which could for instance contribute to control and modify the range of temperature and the magnitude of the MCE.

In order to explore the effect on the electrical transport properties of the presence of LSMO inside the pores of the MSF layer, electric resistance ($R$) was studied as a function of temperature and applied magnetic field. The experiments were performed with $H$ and the electric current ($i$) applied parallel to the substrate of the film. Given the large resistivity of the system, two-points silver paint electrodes were used to apply $i$ = 10 nA, which guaranteed an ohmic response of the system (as determined previously through a current-voltage study within the T working range). Figures 10.a and 10.b display $R(T)$ with $H$ = 0 Oe and $H$ = 30000 Oe for LSMO20/SiO$_2$ and LSMO20/MSF respectively. It can be observed that both samples show an insulating behavior in the whole range of temperatures, which was evaluated until the measured resistance was too high to be reliable with the available equipment ($R > 10^9$ Ω for $T <$ 100 K). Note that LSMO5/MSF and LSMO5/SiO$_2$ were too insulating to be properly characterized while LSMO100/MSF and LSMO100/SiO$_2$ were too thick to show some influence of the MSF layer in this electrodes configuration.

The absence of a metal-insulator transition in $R(T)$ around $T_C$ (Figures 10.a and 10.b) is consistent with the low temperature *FM* insulating phase previously reported for this compound [38], [58]. However, the presence of $H$ induces a reduction in the electric resistance. This magnetoresistance (*MR*) behavior, which increases with the decreasing temperature, is associated with the magnetic disorder at the grain boundaries [59].

Figures 10.c – 10.f compare MR(H) for both samples at different temperatures, showing that their magnitude increases when T decreases, in agreement with Fig. 10.a and 10.b. The MR(H) observed for $T \ll T_C$ is consistent with grain boundary MR effects, already widely reported in polycrystalline samples [59]. However, it can be observed that MR(H) is larger for LSMO20/MSF than LSMO/SiO$_2$ for T > 200 K. Regarding that LSMO5/MSF has the $T_C$ at 290 K (see Table 2), this MR could be associated with the LSMO inside the pores.

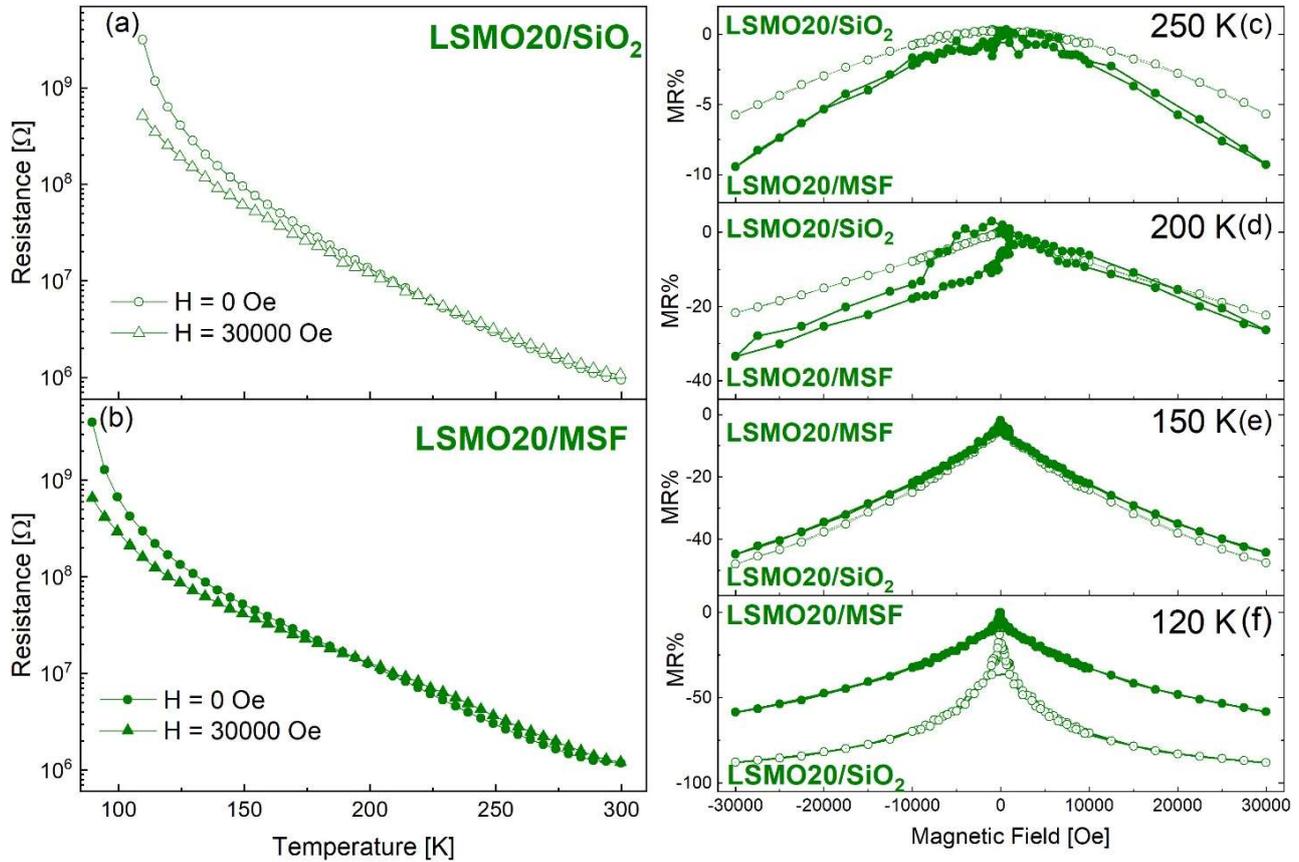

Figure 10: Resistance as a function of temperature measured on cooling with H = 0 T and 3 T, for (a) LSMO20/SiO$_2$ and (b) LSMO20/MSF. Magnetoresistance vs. applied magnetic field for LSMO20/MSF and LSMO20/SiO$_2$ measured at (c) 250 K, (d) 200 K, (e) 150 K and (f) 120 K.

Although the in-plane geometry is not the most favorable one to evaluate the electrical transport properties in multilayers, it was possible to identify a measurable effect due to of the presence of LSMO inside the pores. In order to maximize the effect on the MR in next designs, the current should be applied perpendicular to the plane of the films.

Summarizing, the results presented here have demonstrated a wide range of potentialities for very different device applications. The presence of LSMO within the pores was clearly evidenced in the magnetic properties of the samples. By one hand, the MSF layer seemed to induce an increment of the coercive field even for the thicker LSMO layer sample. By other hand, weakly interacting NPs features appear in M(T) comparing FC and ZFC curves, presenting a blocked state at low temperature. These features become more appreciable when the magnetic contribution of the LSMO top layer diminishes. Thus, mesoporous thin films appear as an interesting template to build ordered ferromagnetic/superparamagnetic assemblies of single domain NPs, with an excellent control of the diameter of the NPs and the distance between them. The high quality of these heterostructures allows the fabrication of multilayers, alternating highly ordered nanoparticle layers templated by mesoporous thin film multilayers with different pore diameter [6]. The electrical properties measured in the LSMO20/MSF system set this work up as a first step to the fabrication of micro and nanodevices based on NPs properties. Definitively, this is an advance for the incorporation of NPs in device applications for spintronic, using either FM metallic or FM insulating compounds [60], [61], as the prototypical $La_{0.67}Sr_{0.33}MnO_3$ or the same $La_{0.88}Sr_{0.12}MnO_3$ explored here respectively. It is expected this could contribute with downscaling for a new generation of spintronic sensors [62]. A simple spintronic valve just controlled by the NPs size is proposed at Figure 11.a. This shows a pillar device fabricated after depositing a single FM metallic manganite (i.e. $La_{0.67}Sr_{0.33}MnO_3$) on a three-layer mesoporous thin film with alternating pore diameter. [6]

Alternatively, we have recently studied mesoporous thin films (MTFs) of Yttria-stabilized Zirconia in order to enhance the interfacial electrochemical properties of SOFCs.[63] Regarding that the use of lanthanum strontium cobaltite perovskites as mixed ionic-electronic conductors (MIEC) are crucial for SOFCs electrodes,[64] [65][66] we propose in Figure 11.b a novel design of SOFCs. The implementation of these architectures are based on the deposition of a MTF of the same electrolyte material, prior to depositing the MIEC electrodes into the nanostructured surface of the electrolyte.

Additionally, the exploration of the magnetocaloric properties promotes these heterostructures as candidates for localized control of the temperature of the mesoporous matrix and/or the inside of the pore, just remotely turning on/off an external magnetic field (See Figure 11.c). The incorporation of these MCE nanostructures in the design of heat exchange devices would avoid an external solid or liquid bath for heating/cooling, reducing the power involved and the thermal inertia.[57] Moreover, the partial filling with remaining accessible porosity found here for LSMO5/MSF exposes the possibility of using this kind of heterostructures with accessible porosity to the environment to produce responsive systems. [67] Furthermore, by combining these features with thermosensitive chemical functionalization of the remaining accessible pores,[68] the system could work for instance as magnetic field-controlled thermal gates.

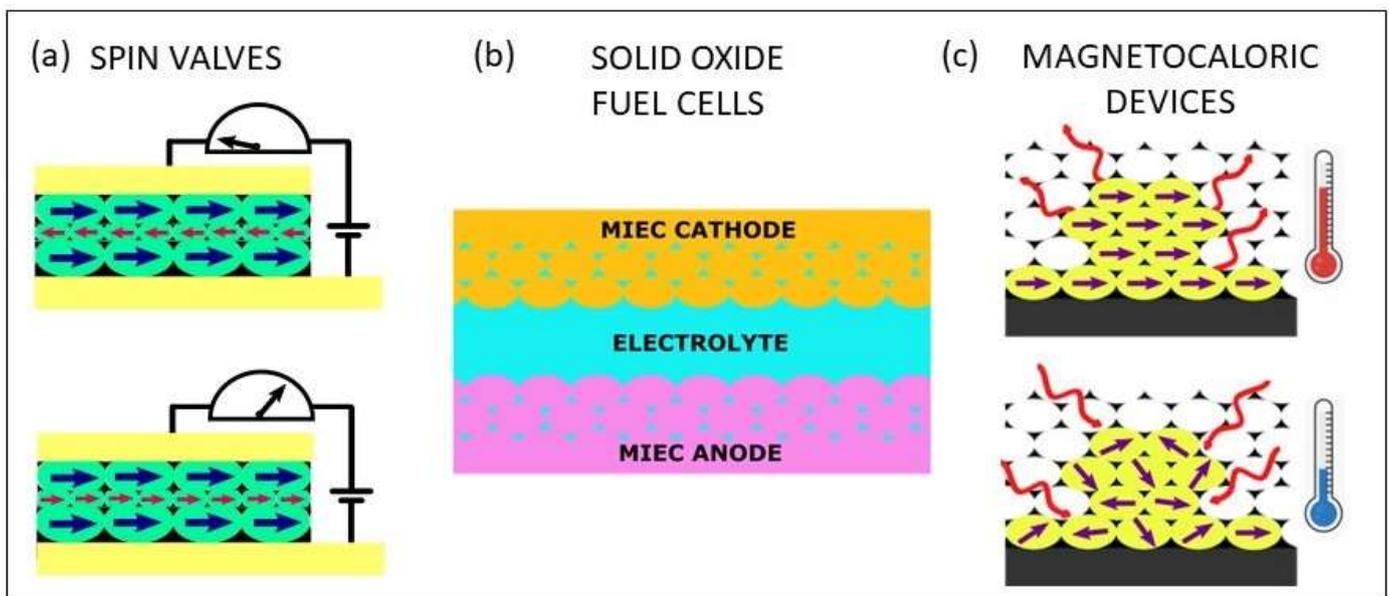

Figure 11: Examples of potential applications for the complex oxide PLD deposition on mesoporous thin films. a) Spintronic devices based on FM metallic manganite nanoparticles. b) Electrolyte – electrode interface optimization in solid oxide fuel cells. c) Magnetic control of heat exchange in MCE devices.

## 3. CONCLUSIONS

In this work, we report that a perovskite complex oxide was successfully synthesized inside the pores of a highly ordered mesoporous thin film by the first time. Our strategy involved the deposition of $La_{0.88}Sr_{0.12}MnO_3$ by pulsed laser ablation onto mesoporous $SiO_2$ films. One of the goals of this method lies in the absence of chemical wastes, avoiding the pores occlusion during the filling process. XRR, SEM, and TEM/EELS structural characterization confirmed the presence of this manganite compound inside the pores and the preservation of the ordered mesoporous structure. Particularly, EELS mapping demonstrated that the LSMO filling is homogenous along the whole thickness of MSF. The filling fraction was analyzed by XRR, revealing that it is possible to get partial filling states with remaining porosity, still accessible to the environment.

The effect of the presence of LSMO nanoparticles within the pores was clearly evidenced in the magnetic, electrical and magnetocaloric properties of the samples. By contrast, equivalent LSMO films deposited on dense $SiO_2$ thin films presented in all the cases the typical properties of polycrystalline LSMO thin films.

Finally, PLD turns out to be a simple and efficient technique to incorporate complex oxide NPs into a mesoporous structure, expanding the actual potentialities of the multifunctional oxides. Therefore, this path could be easily extended to a myriad of materials and applications. Moreover, the high quality of the obtained heterostructures renders them suitable for microfabrication techniques in micro and nanodevices applications.

## EXPERIMENTAL SECTION

*Mesoporous $SiO_2$ films synthesis*: A sol-gel was prepared using well-established procedure, using tetraethoxysilane (TEOS, Merck) as the oxide precursor and Pluronic F127 as a templating agent. The TEOS:F127:$H_2O$:HCl:EtOH molar ratio was 1:0.005:10:0.008:40. The solution was aged for 24 h before film deposition, as it yields more ordered mesoporous structures [5], [48]. It was deposited by dip coating on a commercial Si substrate with 1 µm of dense amorphous thermal oxide ($SiO_2$/Si substrates). After deposition, the film was subjected to the following treatment to improve the miscelles order: 24 h at room temperature and 50% of relative humidity + 24 h at 60 ºC + 24 h at 130 ºC. Then, the temperature was increased at 1 ºC/min up to 350 ºC + 2 h at 350 ºC in order to calcine the surfactant to get the mesoporous structure into the film.

*$La_{0.88}Sr_{0.12}MnO_3$ pulsed laser deposition:* it was deposited using the 266 nm harmonic of an Nd:YAG laser with a pulse frequency of 10 Hz and a fluence of 1 J/cm$^2$. The deposition conditions were 850 °C and 0.1 mbar of $O_2$ pressure. After deposition, the film was cooled down to room temperature under 100 mbar of $O_2$, to reduce the number of oxygen vacancies and to improve its magnetic properties. See Figure 1 for the scheme of LSMOd/MSF sample fabrication.

*Structural characterization:* XRD and XRR measurements were performed using a Panalytical Empyrean diffractometer, to determine the crystalline structure, electronic density, and thickness of the films. In the corresponding cases, XRR measurements were performed in controlled environmental *RH* conditions to determine accessible porosity [47], [45] (A full description of the data analysis is described in the supplementary information SI).

Morphology was thoroughly characterized by scanning and transmission electron microscopy, using SUPRA 40 Zeiss and EM109T Zeiss microscopes respectively.

Cross-sectional samples were prepared in standard 'trench' geometry using focused ion beam (FIB) milling, with Ga$^+$ ions at 30 kV, in a Zeiss Cross Beam 340 microscope. A Pt strap was deposited onto the sample surface to minimize ion damage during subsequent FIB milling. Great care was taken to minimize damage to the lamella by milling with a low beam current parallel to the surface of the final membrane. A Tecnai F20-G2 UT field emission gun (FEG) TEM operating at 200 kV was used to acquire high-resolution lattice images, and a Quantum ER spectrometer was used for elemental mapping and electron energy-loss spectroscopy (EELS). EEL spectra were measured with a collection semiangle of 8 mrad and a dispersion of 0.25 eV/channel. The energy resolution at this condition, estimated from the full width at half maximum of the zero-loss peak, was 1.5 eV. Because of the overlap of Oxygen and Manganese edges, jump-ratio method was used for elemental mapping [69]. The energy-loss images were acquired using an objective aperture defining a collection semiangle of 8 mrad and an energy window width of 15 eV for Mn and 20 eV for La, to provide enough signal for energy-filtered imaging. The images were centered at 625 eV and 644 eV for Mn mapping, and 815 eV and 840 eV for La mapping. The average relative thickness was about 0.4 for the lamella, ensuring that no multiple scattering contributions should be considered.

Magnetic and electrical characterization: $M$ and $R$, as a function of $T$ and $H$, were measured in a Physical Property Measurement System model Versalab (50 K - 400 K and 30000 Oe), manufactured by Quantum Design. In this system, it is possible to measure magnetization through the vibrating sample magnetometer module or the electric transport properties through the ETO module.

## ASSOCIATED CONTENT

### *Supporting Information (SI) available:*

Additional information is supplied about i) XRD, ii) SEM, iii) TEM, iv) TEM analysis, v) XRR and AP analysis, vi) effects of the thermal and atmospherical PLD conditions on the mesoporous silica thin film, vii) complementary magnetization results and viii) additional XRR analysis for the LSMOd/MSF samples.

## AUTHOR INFORMATION

### *Corresponding author*

*leticiagranja@integra.cnea.gob.ar / granja.l@gmail.com

### *Present Address*

†Departamento de Coordinación Proyecto ICES, Centro Atómico Constituyentes, CNEA. Av. General Paz 1499, (B1650KNA) Villa Maipú, Provincia de Buenos Aires, Argentina.

## DATA AVAILABILITY STATEMENT

The data supporting this article have been included as part of the Supplementary Information.

## CREDIT AUTHORSHIP CONTRIBUTION STATEMENT

The manuscript was written through contributions of all authors. / All authors have given approval to the final version.

Sebastián Passanante: Investigation, Methodology, Formal Analysis, Writing-Original draft preparation, Writing- Reviewing and Editing

Mariano Quintero, Andrés Zelcer, Sergio Moreno, Diego Lionello, Daniel Vega: Investigation, Methodology, Formal analysis, Resources, Writing-Reviewing and Editing

Leticia Granja: Investigation, Formal analysis, Supervision, Conceptualization, Methodology, Resources, Writing-Original draft, Writing-Reviewing and Editing, Project administration, Funding acquisition.

**ACKNOWLEDGMENTS**

The authors acknowledge financial support received from ANPCyT (PICT 2021-00495, PICT 2018-02397). We thank M. C. Fuertes and P. C. Angelomé for fruitful discussions; N. D. Mathur for his helpful comments and P. Levy for manuscript reading. We thank to S.J. Ludueña for the SEM images performed at CMA-UBA, Brenda Ledesma (LANAIS-UBA) and Gonzalo Zbihlei (INN-CNEA) for TEM images and cross-section FIB preparation of the samples, performed by Carlos Bertoli (INN-CNEA). All the electronic microscopes used belong to the Sistema Nacional de Microscopía (MINCYT).

**For Table of contents:**

*A successful strategy for the fabrication of complex oxide nanoparticles networks with highly controlled size and ordering is provided. Specifically, the feasibility of synthesizing $La_{0.88}Sr_{0.12}MnO_3$ within the pores of mesoporous $SiO_2$ thin film using pulsed laser deposition is demonstrated. The presence of the nanoparticles is evidenced in the study of the structural, magnetic, magnetocaloric and electrical transport properties of the obtained nanostructures. Their easy incorporation into micro and nanofabrication procedures promises direct implications in the field of interfaces and nanoparticle devices.*

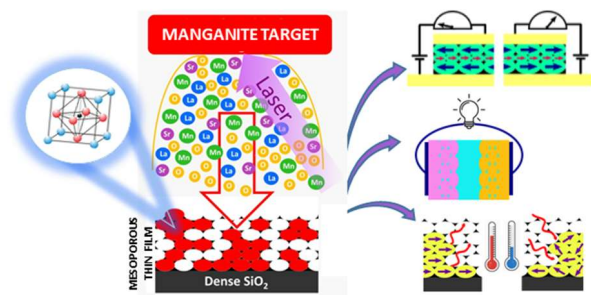